\documentclass[aps,prd,onecolumn,preprintnumbers,notitlepage,superscriptaddress,
nofootinbib,amsfont,amssymb,10pt,singlespacing]{revtex4-1}
\usepackage{amsmath,bm}
\usepackage{times}
\usepackage{braket}
\usepackage{color,graphicx}
\usepackage{slashed}
\usepackage{hyperref}
\usepackage{graphics}
\usepackage{subfigure}
\usepackage{multirow,makecell}
\usepackage{textcomp}
\usepackage{mathtools}
\usepackage{float}

\newcommand{\beq}{\begin{eqnarray}}
\newcommand{\eeq}{\end{eqnarray}}

\def\lsim{ {\ \lower-1.2pt\vbox{\hbox{\rlap{$<$}\lower6pt\vbox{\hbox{$\sim$}
}}}\ } }
\def\gsim{ {\ \lower-1.2pt\vbox{\hbox{\rlap{$>$}\lower6pt\vbox{\hbox{$\sim$}
}}}\ } }
\definecolor{Red}{rgb}{1.,0.,0.}

\definecolor{Blue}{rgb}{0.,0.,1.}

\definecolor{nicered}{rgb}{0.7,0.1,0.1}
\definecolor{nicegreen}{rgb}{0.1,0.5,0.1}

\hypersetup{colorlinks,citecolor=nicegreen,linkcolor=nicered}

\begin{document}

\title{
Study of the four-body decays $B^{0}_{S} \rightarrow \pi\pi\pi\pi$ in the perturbative QCD approach}

\author{Hui-Qin~Liang}
\email[Electronic address:]{2504285029@qq.com}
\affiliation{School of Physical Science and Technology,
 Southwest University, Chongqing 400715, China}

\author{Xian-Qiao~Yu}
\email[Electronic address:]{yuxq@swu.edu.cn}
\affiliation{School of Physical Science and Technology,
Southwest University, Chongqing 400715, China}

\date{\today}

\begin{abstract}

In this work, we analyse the CP-averaged branching ratios and direct CP-violating asymmetries of the four-body decays $B_{S} \rightarrow \pi\pi\pi\pi$ decay from the S-wave resonances, $f_{0}(980)$ and $f_{0}(500)$ and P-wave resonances, $\rho(770)$ by introducing the S-wave and P-wave $\pi\pi$ distribution amplitudes within the framework of the perturbative QCD approach. We also calculate branching ratios of the two-body decays $B^{0}_{S}\rightarrow\rho^{0}\rho^{0}$, $B^{0}_{S}\rightarrow\rho^{+}\rho^{-}$ from the corresponding quasi-two-body decays models and compare our results with those obtained in previous perturbative QCD approach, QCD factorization approach and FAT approach, it is found that the predictions are consistent with present data within errors. The branching ratios of our calculations for the four-body decays $B_{S}\rightarrow \pi\pi\pi\pi$ are at the order of the $10^{-7}$. For the CP-violating asymmetries, we found that CP-violating asymmetry can be enhanced largely by the $\rho-\omega$ mixing resonances when $\pi\pi$ pairs masses are in the vicinity of $\omega$ resonance.

\end{abstract}

\maketitle

%
%

\section{Introduction}\label{sec:intro}

CP-violating asymmetries which associated with weak phase from the CKM matrix and CP-averaged branching ratios have attracted a great deal of attention~\cite{Cabibbo:1963aa,Kobayashi:1973ab}, since they are regard as offering the most important opportunity to testing the standard model and searching new physics beyond standard model. Experimental dates provide several CP violation processes in $K$ and $B$ meson decays processes and frameworks for the two-body decays of $B$ meson with vector and scalar final states have developed in recent decades~\cite{Particle Date Group:2020ac,Chen:2002ad,Zhu:2005ae,Li:2005af,Yan:2018ag}. Comparing with two-body decays, the multi-body decays of $B$ meson are more interesting due to their more complicated processes.

\begin{figure}[htbp]
 \centering
 \begin{tabular}{l}
 \includegraphics[width=0.8\textwidth]{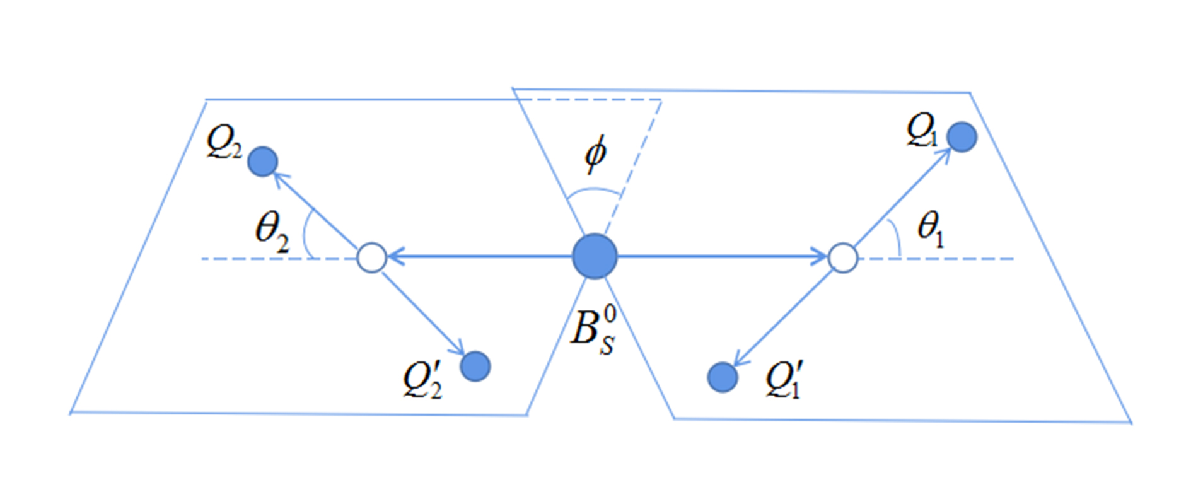}
 \end{tabular}
 \caption{Helicity angles of $(\pi^{+}\pi^{-})(\pi^{+}\pi^{-})$ decays, $\theta$ is defined as the polar angle of $\pi^{+}$ in $\pi^{+}\pi^{-}$ intermediate states and $\phi$ represents the angle among two $\pi\pi$ pairs in the rest frame of $B$ meson.}
   \label{fig:fig1}
 \end{figure}

Experimentally, the four-body decays of $B$ meson with certain two-body invariant mass regions which are shown in Fig.~\ref{fig:fig1} have been collected by LHCb~\cite{Aaij:2012ah,Aaij:2015ai,Aaij:2015aj,Aaij:2015ak,Aaij:2018al,Aaij:2019am,Aaij:2019an}, Belle~\cite{Kyeong:2009ao,Chiang:2010ap}, BaBar~\cite{Aubert:2008aq,Aubert:2008ar} and other collaborations.
Generally, it is not easy to calculate dynamics of these decays, however, it can be simplified by employing the factorization theorems. Several factorization approaches, such as QCD factorizations (QCDF)~\cite{Beneke:2001as,Beneke:2003at,Qi:2020au,Li:2014av,Pavao:2017aw,Akar:2019ax,Qi:2020ay,Huber:2020az,Wang:2021bb}, the soft collinear effective theory (SCET) ~\cite{Bauer:2001bc,Bauer:2006bd,Wang:2017be}, factorization-assisted topological amplitude approach (FAT)~\cite{Wang:2017bf} and the perturbative QCD (PQCD) factorization approachs~\cite{Keum:2001bg,Lu:2001bh,Yu:2010bi,Wang:2014bj,Li:2016bk,Wang:2016bl,Li:2018bm,Ma:2018bn,Ma:2019bo,Li:2019bq,Zou:2020br,Zou:2021bs} are used to investigate these decays.
Comparing with other approaches, the perturbative QCD factorization which based on $K_{T}$ factorization theorem is more appropriate to find out the four-body decays of $B$ meson~\cite{Catani:1990bt,Collins:1991bu,Chen:2004bv}.

In the pQCD factorization framework, we usually use a factorization scale of about 1/b to separate the perturbative area from the non-perturbative area, where b is the conjugate variable obtained by Fourier transformation of the transverse momentum of the quark in the meson. The non-perturbative part below the 1/b energy scale will be included in the wave function that are universal and irrelevant to the process, however, the part above the 1/b energy scale depends on differential decay channels, and the numerical calculations of Feynman diagrams is carried out by using the perturbation theory. For the four-body decay $B^{0}_{S} \rightarrow \pi\pi\pi\pi$, the amplitude can be written as~\cite{Rui:2021bw,Li:2021bx}

\begin{equation}
{\cal A}={\cal H}\otimes \Phi_{B} \otimes \Phi_{h_{1}h_{2}} \otimes \Phi_{h_{3}h_{4}},
\end{equation}

here ${\cal H}$ is hard decay kernel that can be perturbatively calculated, $\Phi_{B}$ is the wave function of $B$ messon, $\Phi_{h_{1}h_{2}}$ and $\Phi_{h_{3}h_{4}}$ are wave functions of $\pi\pi$ pairs, which can be regard as the non-perturbative part and the differential wave functions can be found in the following.

In this work, we focus on the study of the quasi-two-body decays $B^{0}_{S} \rightarrow N_{1}N_{2} \rightarrow \pi\pi\pi\pi$ in which the $\pi\pi$ pair is selected in the low invariant mass range ($<1100 \hspace{0.1cm}{\rm MeV}$)~\cite{Aaij:2015ai} and can arise from S wave and P wave contributions with the vector resonances $\rho^{0}$, $\omega$, the scalar resonances $f_{0}(980)$, $f_{0}(500)$. The strong interactions between S wave, P wave contributions and the final-state pion pairs can not be ignored, so we discuss them by introducing timelike form factors $F_{\pi}$. For the resonances $\rho^{0}$, we adopt the Gounaris-Sakurai model, the BW model is used for resonances $\omega$. $f_{0}(980)$ is parameterized by $ Flatt\acute{e}$ model and $f_{0}(500)$ is modeled by the Breit-Wigner function~\cite{Wang:2015by,Li:2017bz,Zou:2021cd}. The vector or scalar resonances models of the pion pair have been borrowed for the study of quasi-two-body $B$ meson decays and the range of invariant mass in $\pi\pi$ pair varies from 300 MeV to 1100 MeV~\cite{Aaij:2015ai,Li:2021bx}.
Besides, the four-body decays mainly cover six helicity amplitudes $A_{h}$ with $h=VV (3), VS, SV$ and $SS$. The P wave amplitudes in which two pion pair resonances form a vector meson correspond to $h=VV$. For the decays of $B\rightarrow VV$, the amplitude can be defined as three invariant helicity components: $A_{0}$ for longitudinally polarized vector meson and $A_{\parallel}$, $A_{\perp}$ for transversely polarized vector meson~\cite{Huang:2006ce,Li:2006cf}. $h=VS$ and $h=SV$ refer to a S wave or a P wave amplitudes from $N_{1}$ or $N_{2}$ and $h=SS$ is the amplitude that arises from S wave amplitudes in which two pion pair resonances form a scalar meson.

After the introduction above, we shall present theoretical framework in the PQCD factorization approach for four-body decays of $B$ meson in section \ref{sec:pert}, the S-wave and P-wave function of $\pi\pi$ pairs in section \ref{sec:swavef} and section \ref{sec:pwavef}. We parameterize the decay amplitude and direct CP asymmetries of the considered decay modes in section \ref{sec:amp}. In section \ref{sec:numer}, the numerical results and analysis about the two-body and four body decays are collected, and at last, we give a short summary in section \ref{sec:summary}. The factorization formulas for the decay amplitudes are organized into appendix.

%
%
\section{Theoretical Framework}\label{sec:pert}

 For the four-body $B^{0}_{S} \rightarrow \pi\pi\pi\pi$ decay, the weak effective Hamiltonian is given by~\cite{Buchalla:1996cg}

\begin{equation}
{\cal H}_{eff}=\frac{G_{F}}{\sqrt2}\big\{V^*_{ub}V_{uX}[C_{1}(\mu)O_{1}(\mu)+C_{2}(\mu)O_{2}(\mu)]-V^*_{tb}V_{tX}[\sum^{10}_{i=3}C_{i}(\mu)O_{i}(\mu)]\big\}.
\end{equation}

Here $X=(d,s)$, $G_{F}$ is Fermi coupling constant, $V^*_{ub}$$V_{uX}$ and $V^{*}_{tb}$$V_{tX}$ are Cabibbo-Kobayashi-Maskawa (CKM) factors, $C_{i}$ are Wilson coefficients. $O_{i}$ are four-quark operators, and which can be written as:

\begin{equation}
\begin{split}
O_{1}&=\overline{b}_{\alpha}\gamma_{\mu}(1-\gamma_{5})u_{\beta}\overline{u}_{\beta}\gamma^{\mu}(1-\gamma_{5})X_{\alpha},\\
O_{2}&=\overline{b}_{\alpha}\gamma_{\mu}(1-\gamma_{5})u_{\alpha}\overline{u}_{\beta}\gamma^{\mu}(1-\gamma_{5})X_{\beta},\\
O_{3}&=\overline{b}_{\alpha}\gamma_{\mu}(1-\gamma_{5})X_{\alpha}\sum_{X^{'}}\overline{X^{'}}_{\beta}\gamma^{\mu}(1-\gamma_{5})X^{'}_{\beta},\\
O_{4}&=\overline{b}_{\alpha}\gamma_{\mu}(1-\gamma_{5})X_{\beta}\sum_{X^{'}}\overline{X^{'}}_{\beta}\gamma^{\mu}(1-\gamma_{5})X^{'}_{\alpha},\\
O_{5}&=\overline{b}_{\alpha}\gamma_{\mu}(1-\gamma_{5})X_{\alpha}\sum_{X^{'}}\overline{X^{'}}_{\beta}\gamma^{\mu}(1+\gamma_{5})X^{'}_{\beta},\\
O_{6}&=\overline{b}_{\alpha}\gamma_{\mu}(1-\gamma_{5})X_{\beta}\sum_{X^{'}}\overline{X^{'}}_{\beta}\gamma^{\mu}(1+\gamma_{5})X^{'}_{\alpha},\\
O_{7}&=\frac{3}{2}\overline{b}_{\alpha}\gamma_{\mu}(1-\gamma_{5})X_{\alpha}\sum_{X^{'}}e_{X^{'}}\overline{X^{'}}_{\beta}\gamma^{\mu}(1+\gamma_{5})X^{'}_{\beta},\\
O_{8}&=\frac{3}{2}\overline{b}_{\alpha}\gamma_{\mu}(1-\gamma_{5})X_{\beta}\sum_{X^{'}}e_{X^{'}}\overline{X^{'}}_{\beta}\gamma^{\mu}(1+\gamma_{5})X^{'}_{\alpha},\\
O_{9}&=\frac{3}{2}\overline{b}_{\alpha}\gamma_{\mu}(1-\gamma_{5})X_{\alpha}\sum_{X^{'}}e_{X^{'}}\overline{X^{'}}_{\beta}\gamma^{\mu}(1-\gamma_{5})X^{'}_{\beta},\\
O_{10}&=\frac{3}{2}\overline{b}_{\alpha}\gamma_{\mu}(1-\gamma_{5})X_{\beta}\sum_{X^{'}}e_{X^{'}}\overline{X^{'}}_{\beta}\gamma^{\mu}(1-\gamma_{5})X^{'}_{\alpha},
\end{split}
\end{equation}
where $\alpha$ and $\beta$ are color indices, ${X^{'}}=u,d,s,c$ or $b$ quarks. $O_{1}$ and $O_{2}$ are tree operators, $O_{i}(i=3,...,10)$ are penguin operators, in which $O_{i}(i=7,...,10)$ are the operators from electroweak penguin diagrams.

The light-cone coordinate system is used in the B meson rest frame, and the system is expressed as

\begin{equation}
\begin{split}
p^{+}=\frac{p^{0}+p^{3}}{\sqrt{2}},\qquad
p^{-}=\frac{p^{0}-p^{3}}{\sqrt{2}},\qquad
p_{\top}=(p^{1},p^{2}),
\end{split}
\end{equation}

through the following relational formula

\begin{equation}
\begin{split}
p^{2}={2}p^{+}p^{-}-p^{2}_{\top},\qquad
p_{^{1}}\cdot p_{^{2}}=p^{+}_{1}p^{-}_{2}+p^{-}_{1}p^{+}_{2}-p_{{1}{\top}}\cdot p_{{2}{\top}}.
\end{split}
\end{equation}

\begin{figure}[htbp]
 \centering
 \begin{tabular}{l}
 \includegraphics[width=0.8\textwidth]{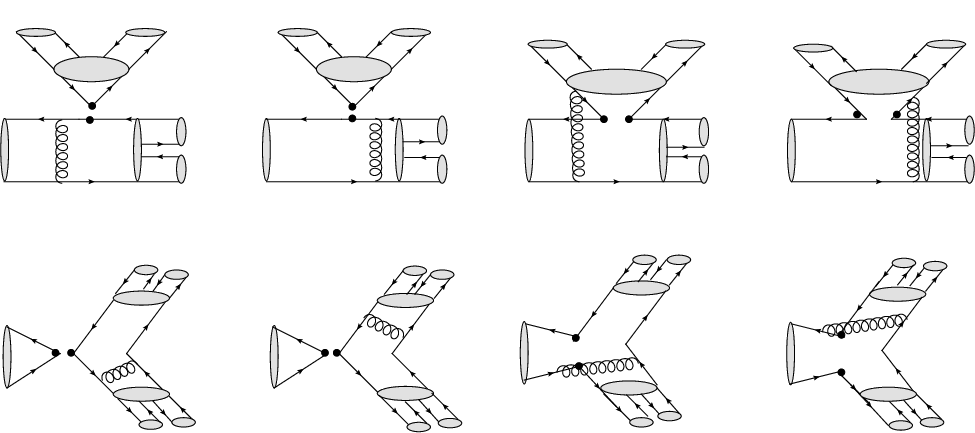}
 \end{tabular}
 \caption{The lowest order Feynman diagrams for the $B_{S} \rightarrow N_{1}N_{2} \rightarrow \pi\pi\pi\pi$ decays}
   \label{fig:fig2}
 \end{figure}

For the $B(p_{B}) \rightarrow N_{1}(p)N_{2}(q) \rightarrow Q_{1}(q_{1})Q^{'}_{1}(q^{'}_{1})Q_{2}(q_{2})Q^{'}_{2}(q^{'}_{2})$ decay, we choose the B meson mass $M_{B}$, $p_{B}=p+q$, $p=q_{1}+q^{'}_{1}$, $q=q_{2}+q^{'}_{2}$ and let $N_{1}$ and $N_{2}$ intermediate states move along with the direction of $n=(1,0,0_{\top})$ and $v=(0,1,0_{\top})$ respectively, and the Feynman diagrams have been described in Fig.~\ref{fig:fig2}. So we define the intermediate states and quark momentum as~\cite{Rui:2021bw,Li:2021bx,Zhang:2021ch}
\begin{equation}
\begin{split}
p_{B}&=\frac{M_{B}}{\sqrt{2}}(1,1,0_{\top}),        \qquad                                                           k_{B}=(0,\frac{M_{B}}{\sqrt{2}}{x_{B}},k_{B\top}),\\
p&=\frac{M_{B}}{\sqrt{2}}(g^{+},g^{-},0_{\top}),  \qquad                                                             k_{p}=(\frac{M_{B}}{\sqrt{2}}{x_{1}}g^{+},0,k_{1\top}),\\
q&=\frac{M_{B}}{\sqrt{2}}(f^{-},f^{+},0_{\top}),  \qquad                                                             k_{q}=(0,\frac{M_{B}}{\sqrt{2}}x_{2}f^{+},k_{2\top}).
\end{split}
\end{equation}

The above factors are
\begin{equation}
\begin{split}
g^{\pm}=\frac{1}{2}[1+\eta_{1}-\eta_{2}\pm{\sqrt{(1+\eta_{1}-\eta_{2})^{2}-{4}\eta_{1}}}],\\
f^{\pm}=\frac{1}{2}[1-\eta_{1}+\eta_{2}\pm{\sqrt{(1+\eta_{1}-\eta_{2})^{2}-{4}\eta_{1}}}],
\end{split}
\end{equation}

where $\eta_{1,2}=\omega^2_{1,2}/M^2$ are mass ratios, and the invariant mass $\omega^2_{1,2}$ and their momentum $p$,$q$ satisfy the relation $\omega^2_{1}=p^2$ and $\omega^2_{2}=q^2$. $\emph{x}_{i}$, $i=B,{1},{2}$ indicate momentum fractions inside the meson and they run between $0\sim1$. We also study $P$ wave pairs by introducing corresponding longitudinal polarization vectors, and the vectors can be written as
\begin{equation}
\begin{split}
\epsilon_{p}=\frac{1}{\sqrt{2\eta_{1}}}(g^{+},-g^{-},0_{\top}) , \qquad                           \epsilon_{q}=\frac{1}{\sqrt{2\eta_{2}}}(-f^{-},f^{+},0_{\top}),
\end{split}
\end{equation}

with $\epsilon^{2}_{p}=\epsilon^{2}_{q}={-1}$ and $\epsilon_{p}\cdot {p}=\epsilon_{q}\cdot {q}={0}$.

Considering the final state meson $q_{1},q^{'}_{1}$ and $q_{2},q^{'}_{2}$, we decompose them as

\begin{equation}
\begin{split}
q_{1}&=(\frac{M_{B}}{\sqrt{2}}g^{+}(\zeta_{1}+\frac{(r_{1}-r^{'}_{1})}{2\eta_{1}}),\qquad
\frac{M_{B}}{\sqrt{2}}g^{-}({1}-\zeta_{1}+\frac{(r_{1}-r^{'}_{1})}{2\eta_{1}}),p_{\top}),\\                           q^{'}_{1}&=(\frac{M_{B}}{\sqrt{2}}g^{+}(1-\zeta_{1}-\frac{(r_{1}-r^{'}_{1})}{2\eta_{1}}),\qquad
\frac{M_{B}}{\sqrt{2}}g^{-}(\zeta_{1}-\frac{(r_{1}-r^{'}_{1})}{2\eta_{1}}),-p_{\top}),\\
q_{2}&=(\frac{M_{B}}{\sqrt{2}}f^{-}(1-\zeta_{2}+\frac{(r_{2}-r^{'}_{2})}{2\eta_{2}}),\qquad
\frac{M_{B}}{\sqrt{2}}f^{+}(\zeta_{2}+\frac{(r_{2}-r^{'}_{2})}{2\eta_{2}}),q_{\top}),\\
q^{'}_{2}&=(\frac{M_{B}}{\sqrt{2}}f^{-}(\zeta_{2}-\frac{(r_{2}-r^{'}_{2})}{2\eta_{2}}),\qquad
\frac{M_{B}}{\sqrt{2}}f^{+}(1-\zeta_{2}-\frac{(r_{2}-r^{'}_{2})}{2\eta_{2}}),-q_{\top}),\\
\end{split}
\end{equation}

here the mass ratios $r_{i}=\frac{{M^2_i}}{M^2_B}$,$r^{'}_{i}=\frac{M^{(')2}_{i}}{M^2_B}$. By introducing variables $\zeta_{i}$ $(i=1,2)$, we can derived the meson momentum fractions

\begin{equation}
\begin{split}
\frac{q^{+}_{1}}{p^+}=\zeta_{1}+\frac{(r_{1}-r^{'}_{1})}{2\eta_{1}} ,  \qquad                         \frac{q^{-}_{2}}{q^-}=\zeta_{2}+\frac{(r_{2}-r^{'}_{2})}{2\eta_{2}}.
\end{split}
\end{equation}

The transverse momenta are given by

\begin{equation}
\begin{split}
p^{2}_{\top}=\zeta_{1}(1-\zeta_{1})\omega^{2}_{1}+\frac{(m^{2}_{1}-m^{(')2}_{1})^{2}}{4\omega^{2}_{1}}-\frac{(m^{2}_{1}+m^{(')2}_{1})}{2}, \\                          q^{2}_{\top}=\zeta_{2}(1-\zeta_{2})\omega^{2}_{2}+\frac{(m^{2}_{2}-m^{(')2}_{2})^{2}}{4\omega^{2}_{2}}-\frac{(m^{2}_{2}+m^{(')2}_{2})}{2}.
\end{split}
\end{equation}

One can make the above formula simple by introducing

\begin{equation}
\begin{split}
\alpha_{i}=\frac{(r_{i}-r^{'}_{i})^{2}}{4\eta_{i}}-\frac{(r_{i}+r^{'}_{i})}{2\eta_{i}}.
\end{split}
\end{equation}

Then we can deduce the following relationships for $\zeta_{i}$ and polar angle $\theta_{i}$~\cite{Hambrock:2016ci}.
\begin{equation}
\begin{split}
{2}\zeta_{i}-{1}=\sqrt{1+4\alpha_{i}}\cos\theta_{i},
\end{split}
\end{equation}

\begin{equation}
\begin{split}
\zeta_{i}\in[\frac{1-\sqrt{1+4\alpha_{i}}}{2},\frac{1+\sqrt{1+4\alpha_{i}}}{2}].
\end{split}
\end{equation}

In the PQCD approach the wave functions are treated as nonperturbative inputs. For $B_{X}$ $(X=u,s,d)$, the wave function can be expressed as~\cite{Beneke:2000cj,Kawamura:2001ck}.

\begin{equation}
\Phi_{B}=\frac{i}{\sqrt{2N_{c}}}(\not {p}_{B}+M_{B}){\gamma_{5}}{\phi_{B}({x_{B},b_{B}})},
\end{equation}
where $N_{c}={3}$ is the number of colors, and the distribution amplitude $\phi_{B}$ can be chosen as~\cite{Kurimoto:2002cl,Li:2003cn}
\begin{equation}
\phi_{B}(x_{B},b_{B})=\emph{N}_{B}{{x_{B}}^2}(1-{x_{B}})^2\exp[-\frac{M^2_{B}{{x_{B}}^2}}{2\omega^2_{B}}-\frac{1}{2}(\omega_{B}{b_{B}})^2],
\end{equation}
with the normalization
\begin{equation}
\int^{1}_{0}dx\phi_{B}(x,b=0)=\frac{f_{B}}{{2}\sqrt{2N_{c}}},
\end{equation}
where $N_{B}=91.784$GeV is the normalization constant, $f_{B}$ is the decay constant. For $B^{0}_{s}$ meson, we use the shape parameter $\omega_{B_{s}}=0.48\pm0.048$GeV ~\cite{Ali:2007co}.

At the same time, for the two-meson DAs, we will discuss the S-wave and P-wave via intermediate resonances $f_{0}(980)$, $f_{0}(500)$, $\rho(770)$ and $\omega(782)$, which are listed in Table \ref{1}. The corresponding time-like form factors of them are collected below~\cite{Li:2021bx}.

\begin{table}[htbp]
\centering
\caption{The widths, masses and decay models of intermediate states in our framework.}
\label{1}
\begin{tabular*}{\columnwidth}{@{\extracolsep{\fill}}lllll@{}}
\hline
\hline
$Resonance$            & $Mass[Me V]$          & $Width[Me V]$             & $Model$       & $J^{P}$   \\
\hline
  \\
$f_{0}(980)$           & $ 990\pm20$      & $ 65\pm45$    & $ Flatt\acute{e}$     & $ 0^{+}$ \\
\\
$f_{0}(500)$           & $ 471\pm21$      & $ 534\pm53$      & $BW$        & $ 0^{+}$           \\
\\
$\rho(770)$        & $ 775.26\pm0.25$      & $ 149.1\pm0.8$      & $GS$        & $ 1^{-}$           \\
\\
$\omega(782)$        & $ 782.65\pm0.12$      & $ 8.49\pm0.08$      & $BW$        & $ 1^{-}$           \\
\hline
\hline
\end{tabular*}
\end{table}

\section{S Wave Function}\label{sec:swavef}

For the quasi-two-body decays $B^{0}_{S} \rightarrow N_{1}N_{2} \rightarrow \pi\pi\pi\pi$, we proceed it mainly via quasi-two-body channels, which contain S wave and P wave pion-pair resonant state. Similar to previous Ref~\cite{Zou:2021cd}, S-wave two-pion distribution amplitudes are written as
\begin{equation}
\Phi^{S}_{\pi\pi}=\frac{1}{\sqrt{2 N_{c}}}[\not {p}\phi^{0}_{S}(x,\omega)+\omega\phi^{s}_{S}(x,\omega)+\omega(\not{n}\not {v}-1)\phi^{t}_{S}(x,\omega)],
\end{equation}
with the Gegenbauer coefficient $a_{S}$ and the twist-2 and twist-3 light-cone distribution amplitudes $\phi^{0}_{S}(x,\omega)$, $\phi^{s}_{S}(x,\omega)$, $\phi^{t}_{S}(x,\omega)$.
\begin{equation}
\begin{split}
\phi^{0}_{S}(x,\omega)&=\frac{9F_{S}(\omega^{2})}{\sqrt{2 N_{c}}}a_{S}x(1-x)(1-2x),                                                               \\
\phi^{s}_{S}(x,\omega)&=\frac{F_{S}(\omega^{2})}{{2}\sqrt{2 N_{c}}},                                                                                   \\
\phi^{t}_{S}(x,\omega)&=\frac{F_{S}(\omega^{2})}{{2}\sqrt{2 N_{c}}}(1-2x).
\end{split}
\end{equation}

$F_{S}(\omega^{2})$ is time-like form factor. For a narrow resonance, we consider Breit-Wigner line shape to describe, such as $f_{0}(500)$, the $d\bar{d}$ component in the S-wave amplitude

\begin{equation}
F_{S}(\omega^{2})=\frac{cm^{2}_{f_{0}(500)}}{m^{2}_{f_{0}(500)}-\omega^{2}-{i}m_{f_{0}(500)}\Gamma_{f_{0}(500)}(\omega^{2})},
\end{equation}
with
\begin{equation}
\Gamma_{S}(\omega^{2})=\Gamma_{S}\frac{m}{\omega}{\sqrt{\frac{\omega^2-4m^{2}_{\pi}}{m^{2}_{S}-4m^{2}_{\pi}}}}F^{2}_{R}.
\end{equation}

For the resonance $f_{0}(980)$, although the time-like scalar form factor of $f_{0}(980)$, $f_{0}(1500)$, $f_{0}(1790)$ are $s\bar{s}$ component, different from the last two resonances, the mass of $KK$ system in $f_{0}(980)$ is around 0.98 GeV. So we replace BW formula with $ Flatt\acute{e}$ model, which has motivated by Refs~\cite{Wang:2015by,Liang:2019cp} and works well.

\begin{equation}
\begin{split}
F_{S}(\omega^{2})=\frac{c_{1}m^{2}_{f_{0}(980)}e^{i\theta_{1}}}{m^{2}_{f_{0}(980)}-\omega^{2}-{i}m_{f_{0}(980)}(g_{\pi\pi}\rho_{\pi\pi}+g_{KK}\rho_{KK})}\\
+\frac{c_{2}m^{2}_{f_{0}(1500)}e^{i\theta_{2}}}{m^{2}_{f_{0}(1500)}-\omega^{2}-{i}m_{f_{0}(1500)}\Gamma_{f_{0}(1500)}(\omega^{2})}\\
+\frac{c_{3}m^{2}_{f_{0}(1790)}e^{i\theta_{3}}}{m^{2}_{f_{0}(1790)}-\omega^{2}-{i}m_{f_{0}(1790)}\Gamma_{f_{0}(1790)}(\omega^{2})}.
\end{split}
\end{equation}

where $c_{i}$ and $\theta_{i}$ ($i=1,2,3$) are tunable parameters. $c_{1}=0.9$, $c_{2}=0.106$, $c_{3}=0.066$. $g_{\pi\pi}=0.167$ GeV and $g_{KK}=3.47g_{\pi\pi}$ are coupling constants~\cite{Aaij:2013cq,Aaij:2014cr,Back:2018cs,Xing:2019ct}. $\rho_{\pi\pi}$ and $\rho_{KK}$ are phase space that can be expressed as

\begin{equation}
\begin{split}
\rho_{\pi\pi}=\frac{2}{3}\sqrt{1-\frac{4m^2_{\pi^\pm}}{\omega^2}}+\frac{1}{3}\sqrt{1-\frac{4m^2_{\pi^0}}{\omega^2}},\\
\rho_{KK}=\frac{1}{2}\sqrt{1-\frac{4m^2_{K^\pm}}{\omega^2}}+\frac{1}{2}\sqrt{1-\frac{4m^2_{K^0}}{\omega^2}}.
\end{split}
\end{equation}

\section{P Wave Function}\label{sec:pwavef}

 The P wave resonant states are associated with longitudinal and transverse polarizations. The relevant DAs and time-like scalar form factor can be got from Ref.~\cite{Rui:2018cu}

\begin{equation}
\begin{split}
\Phi^{L}_{P(\pi\pi)}&=\frac{1}{\sqrt{2 N_{c}}}[\omega\not {\epsilon_{p}}\phi^{0}_{P}(x,\omega)+\omega\phi^{s}_{P}(x,\omega)+\frac{\not{p_{1}}\not {p_{2}}-\not{p_{2}}\not {p_{1}}}{\omega(2\zeta-1)}\phi^{t}_{P}(x,\omega)](2\zeta-1),\\
\Phi^{T}_{P(\pi\pi)}&=\frac{1}{\sqrt{2 N_{c}}}[\gamma_{5}\not {\epsilon_{T}}\not {p}\phi^{T}_{P}(x,\omega)+\omega\gamma_{5}\not {\epsilon_{T}}\phi^{a}_{P}(x,\omega)+i\omega\frac{\epsilon^{\mu\nu\rho\sigma}\gamma_{\mu}\epsilon_{Tv}p_{\rho}n_{-\sigma}}{{p}\cdot{n-}}\phi^{v}_{P}(x,\omega)]{\sqrt{\zeta(1-\zeta)+\alpha}},
\end{split}
\end{equation}

with
\begin{equation}
\begin{split}
\phi^{0}_{P}(x,\omega)&=\frac{3F^{\parallel}_{P}(\omega^{2})}{\sqrt{2 N_{c}}}x(1-x)[1+a^{0}_{V}\frac{3}{2}(5(1-2x)^{2}-1)],                                                               \\
\phi^{s}_{P}(x,\omega)&=\frac{3F^{\perp}_{P}(\omega^{2})}{{2}\sqrt{2 N_{c}}}(1-2x)[1+a^{s}_{V}(10x^{2}-10x+1)],                                                                                   \\
\phi^{t}_{P}(x,\omega)&=\frac{3F^{\perp}_{P}(\omega^{2})}{{2}\sqrt{2 N_{c}}}(1-2x)^{2}[1+a^{t}_{V}\frac{3}{2}(5(1-2x)^{2}-1)],                   \\
\phi^{T}_{P}(x,\omega)&=\frac{3F^{\perp}_{P}(\omega^{2})}{{2}\sqrt{2 N_{c}}}x(1-x)[1+a^{T}_{V}\frac{3}{2}(5(1-2x)^{2}-1)],                   \\
\phi^{a}_{P}(x,\omega)&=\frac{3F^{\parallel}_{P}(\omega^{2})}{{4}\sqrt{2 N_{c}}}(1-2x)[1+a^{a}_{V}(10x^{2}-10x+1)],                                                                               \\
\phi^{v}_{P}(x,\omega)&=\frac{3F^{\parallel}_{P}(\omega^{2})}{{8}\sqrt{2 N_{c}}}{[1+(1-2x)^{2}]+a^{v}_{V}[3(2x-1)^{2}-1]}.
\end{split}
\end{equation}

Here $\phi^{0}_{P}(x,\omega)$ and $\phi^{T}_{P}(x,\omega)$ are twist-2 DAs, $\phi^{s}_{P}(x,\omega)$, $\phi^{t}_{P}(x,\omega)$, $\phi^{a}_{P}(x,\omega)$ and $\phi^{v}_{P}(x,\omega)$ are twist-3 DAs, which are associated with the longitudinal and transverse polarization. $a^{0,s,t}_{V}$ and $a^{T,a,v}_{V}$ are Gegenbauer moments, which are determined in Ref~\cite{Li:2021bx}. For the time-like factors $F^{\parallel}_{P}$ and $F^{\perp}_{P}$, we postulate the approximation $F^{\perp}_{P}=(f^{T}_{V}/f_{V})F^{\parallel}_{P}$.

We take the $\rho-\omega$ interference and the excited states into account for the form factor,

\begin{equation}
F^{\parallel}(\omega^{2})=[GS_{\rho}(s,m_{\rho},\Gamma_{\rho})\frac{{1}+c_{\omega}BW_{\omega}(s,m_{\omega},\Gamma_{\omega})}{1+c_{\omega}}+\sum_{i} c_{i}GS_{i}(s_{i},m_{i},\Gamma_{i})][1+\sum_{i} c_{i}]^{-1},
\end{equation}
where $i=\rho(1450), \rho(1700), \rho(2254)$ and $s=m^{2}(\pi\pi)$ is the pion-pair invariant mass square. For the $\omega$ resonant state, we adopt the BW model, however, for $\rho$ resonant state, the GS model based on the BW model is used. These models can be found in Refs.~\cite{Ma:2018bn,Li:2017bz,Lee:2012cv}.

\begin{equation}
GS_{\rho}(s,m_{\rho},\Gamma_{\rho})=\frac{m^{2}_{\rho}[1+d(m_{\rho})\Gamma_{\rho}/m_{\rho}]}{m^{2}_{\rho}-s+f(s,m_{\rho},\Gamma_{\rho})-im_{\rho}\Gamma(s,m_{\rho},\Gamma_{\rho})},
\end{equation}
where
\begin{equation}
\begin{split}
\Gamma(s,m_{\rho},\Gamma_{\rho})&=\Gamma_{\rho}\frac{s}{m^{2}_{\rho}}(\frac{\beta_{\pi}(s)}{\beta_{\pi}(m^{2}_{\rho})})^{3},\\
d{m}&=\frac{3}{\pi}\frac{m^{2}_{\pi}}{k^{2}(m^{2})}\ln(\frac{m+2k(m^{2})}{2m_{\pi}})+\frac{m}{2{\pi}k(m^{2})}-\frac{m^{2}_{\pi}m}{{\pi}k^{3}(m^{2})},\\
f(s,m,\Gamma)&=\frac{\Gamma m^{2}}{k^{3}(m^{2})}[k^{2}(s)[h(s)-h(m^{2})]+(m^{2}-s)k^{2}(m^{2})h^{'}(m^{2})],\\
\end{split}
\end{equation}
with
\begin{equation}
\begin{split}
k(s)&=\frac{1}{2}\sqrt{s}\beta_{\pi}(s),\\
h(s)&=\frac{2}{\pi}\frac{k(s)}{\sqrt{s}}\ln(\frac{\sqrt{s}+2k(s)}{2m_{\pi}}),\\
\beta_{\pi}(s)&=\sqrt{1-4\frac{m^{2}_{\pi}}{s}}.
\end{split}
\end{equation}

\section{The decay Amplitudes}\label{sec:amp}

The rate for the $B^{0}_{S} \rightarrow N_{1}N_{2} \rightarrow \pi\pi\pi\pi$ decay in the $B^{0}_{S}$ meson rest frame can be described as~\cite{Geng:2012cw,Cheng:2018cx,Hsiao:2017cy}
\begin{equation}\label{equ:dbr}
\frac{d\cal{B}}{d\Omega}=\frac{\tau_{B^{0}_{S}}k(\omega_{1}){k(\omega_{2})}{k(\omega_{1},\omega_{2})}}{16(2\pi)^{6} M_{B^{0}_{S}}^2} \mid {\cal A} \mid^2,
\end{equation}

with $k(\omega)=\frac{\sqrt{\lambda(\omega^{2},m^{2}_{h_{1}},m^{2}_{h_{2}})}}{2\omega}$ and $k(\omega_{1},\omega_{2})=\frac{\sqrt{[M^{2}_{B}-(\omega_{1}+\omega_{2})^{2}][M^{2}_{B}-(\omega_{1}-\omega_{2})^{2}]}}{2M_{B}}$ in the pion-pair center-of-mass system. $\tau_{B^0_{s}}$ is lifetime, $\Omega$ stands for $\theta_{1},\theta_{2},\phi,\omega_{1},\omega_{2}$, the K\"{a}ll\'{e}n function $\lambda(a,b,c)= a^{2}+b^{2}+c^{2}-2(ab+ac+bc)$.

The six helicity amplitudes are involved in four body decay of $B$ meson and the relation between total amplitude and other components can be found in Refs~\cite{Rui:2021bw,Li:2021bx,Zhang:2021ch}. Then branching ratio from Eq.~(\ref{equ:dbr}) is replaced by follows

\begin{equation}
{\cal B}_{h}=\frac{\tau_{B}}{4(2\pi)^{6}m^{2}_{B}}\frac{2\pi}{9}C_{h}\int d\omega_{1}d\omega_{2}k(\omega_{1})k(\omega_{2})k(\omega_{1},\omega_{2})\mid A_{h}\mid^{2},
\end{equation}

with
\begin{eqnarray*}
C_{h} =
\left\{
  \begin{array}{lll}
  (1+4\alpha_{1})(1+4\alpha_{2}) \qquad  &h=0,\parallel,\perp\\
  3(1+4\alpha_{1,2})                                             &h=VS,SV\\
  9                                                            &h=SS.
  \end{array}
\right.
\end{eqnarray*}

Here integrations over $\zeta_{1}$, $\zeta_{2}$ and $\phi$ is defined in $C_{h}$.

For the CP-averaged branching ratio, we adopt the same definition as in the Ref~\cite{Zhu:2005ae}. $\overline{{\cal B}}_{h}$ is branching ratio of the charge conjugate channel for $B^{0}_{S} \rightarrow N_{1}N_{2} \rightarrow \pi\pi\pi\pi$ and other expressions can be found in Refs~\cite{Liang:2020cz,Li:2021dc}.

We define

\begin{equation}\label{equ:cde}
{\cal B}^{avg}_{h}=\frac{1}{2}({\cal B}_{h}+\overline{{\cal B}}_{h}),
\end{equation}

and $f_{0,\parallel,\perp}$ is the polarization fraction corresponding P-wave amplitudes,
\begin{equation}
f_{0,\parallel,\perp}=\frac{{\cal B}_{0,\parallel,\perp}}{{\cal B}_{total}},
\end{equation}

with ${\cal B}_{total}={\cal B}_{0}+{\cal B}_{\parallel}+{\cal B}_{\perp}$.

\section{Numerical Results And Discussions}\label{sec:numer}

Based on the framework above, we start our calculations by introducing input parameters, which are listed in Table \ref{1111}, covering the mass of the involved mesons(in GeV), the lifetime of $B_{s}$ meson, the decay constants of $B^0_{s}$, $\rho$, $\omega$ mesons and Wolfenstein parameters\cite{Particle Date Group:2020ac}. The Gegenbauer moments are listed in Table \ref{2222}~\cite{Li:2021bx}.

\begin{table}[htbp]
\centering
\caption{The input parameters of the $B^{0}_{S} \rightarrow N_{1}N_{2} \rightarrow \pi\pi\pi\pi$ decay}
\label{1111}
\begin{tabular*}{\columnwidth}{@{\extracolsep{\fill}}lllll@{}}
\hline
\hline
\\
Masses of the involved mesons     &$M_{B^{0}_{s}}=5.367$ {\rm GeV}       &$M_{\pi^{\pm}}=0.140$ {\rm GeV}   \\
                                  \\
                                  &$m_{b}=4.8$ {\rm GeV}                   &$m_{c}=1.27$ {\rm GeV}               &$m_{\pi^{0}}=0.135$ {\rm GeV}   \\
                                  \\
                                  &$m_{f_{0}(980)}=0.99\pm0.02$ {\rm GeV}            &$m_{f_{0}(500)}=0.50$ {\rm GeV}                \\
                                  \\
                                  &$m_{\rho(770)}=0.775\pm0.02$ {\rm GeV}            &$m_{\omega(782)}=0.78265$ {\rm GeV}                   \\
                                  \\
Decay widths                     &$\Gamma_{\rho^{0}(770)}=0.1491$ {\rm GeV}          &$\Gamma_{\omega(782)}=8.49\times10^{-3}${\rm GeV}    \\
                                  \\
Decay constants                   &$f_{B^0_{s}}=0.24 \pm 0.02$ {\rm GeV}  &$f_{\rho}=0.216 \pm 0.003$ {\rm GeV}  &$ f^{T}_{\rho}=0.184 $ {\rm GeV}  \\
                                   \\
                                    &$f_{\omega}=0.187 \pm 0.005$ {\rm GeV} &$f^{T}_{\omega}=0.151 \pm 0.009$ {\rm GeV}\\
                                   \\
Lifetime of meson               &$\tau_{B^{0}_{s}}=1.512$ {\rm ps}    \\
                                   \\
Wolfenstein parameters           &$\lambda=0.22650$            &$\emph{A}=0.790$\\
                                   \\
                                   &$\bar{\rho}=0.141$   &$\bar{\eta}=0.357$\\
                                   \\
\hline
\hline
\end{tabular*}
\end{table}

\begin{table}[htbp]
\centering
\caption{The Gegenbauer moments are collected from Ref.~\cite{Li:2021bx}.}
\label{2222}
\begin{tabular*}{\columnwidth}{@{\extracolsep{\fill}}lllll@{}}
\hline
\hline
$a_{S}=0.2\pm0.2$            & $a^{0}_{\rho}=0.08\pm0.13$          & $a^{s}_{\rho}=-0.23\pm0.24$      & $a^{t}_{\rho}=-0.354\pm0.062$               \\
  \\
$a^{T}_{\rho}=0.50\pm0.50$           & $ a^{a}_{\rho}=0.40\pm0.40$      & $a^{v}_{\rho}=-0.50\pm0.50$         \\
\hline
\hline
\end{tabular*}
\end{table}

In Table \ref{4444}, we present the CP-averaged branching ratios of the $B^{0}_{S} \rightarrow V_{1}V_{2} \rightarrow \pi\pi\pi\pi$ decays in PQCD approach(here $V$ stands for the vector resonance). The first main uncertainty of these results comes from the QCD scale $\Lambda_{QCD}=0.25\pm0.05$ GeV, the second error from hard scale $t$, which varies from $0.75t\sim1.25t$, and the third error from the Gegenbauer moments $a^{0}$, $a^{s}$, $a^{t}$ and the moments $a^{T}$, $a^{a}$,$a^{v}$ in transversely polarized wave functions. Other errors such as the decay constants of the $B^{0}_{S}$ and the Wolfenstein parameters, are small and can be neglected. From Table \ref{4444}, we can see that the prediction results for branching ratios of pure annihilation decays $B^{0}_{S}\rightarrow\rho^{0}\rho^{0}\rightarrow \pi\pi\pi\pi$ and $B^{0}_{S}\rightarrow\rho^{+}\rho^{-} \rightarrow \pi\pi\pi\pi$, which all cover two kinds of topological penguin diagrams contributions, are at the order of $10^{-8}$ with large uncertainties.

\begin{table}[htbp]
\centering
\caption{The CP-averaged branching ratios of the $B^{0}_{S} \rightarrow V_{1}V_{2} \rightarrow \pi\pi\pi\pi$ decay (in unit of $10^{-8}$), the errors come from QCD scale, hard scale and the Gegenbauer moments.}
\label{4444}
\begin{tabular*}{\columnwidth}{@{\extracolsep{\fill}}lllll@{}}
\hline
\hline
Components  &$B^{0}_{S}\rightarrow(\rho^{+}\rightarrow)\pi^{+}\pi^{0}(\rho^{-}\rightarrow)\pi^{-}\pi^{0}$ &$B^{0}_{S}\rightarrow(\rho^{0}\rightarrow)\pi^{+}\pi^{-}(\rho^{0}\rightarrow)\pi^{+}\pi^{-}$ \\
\hline
\\
$B_{0}$           &$0.76^{+0.30+0.26+0.07}_{-0.21-0.20-0.03}$           &$0.38^{+0.12+0.12+0.01}_{-0.10-0.12-0.01}$            \\
                                  \\
$B_{\parallel}$           &$0.04^{+0.03+0.02+0.02}_{-0.02-0.01-0.00}$          &$0.02^{+0.01+0.01+0.01}_{-0.01-0.01-0.00}$            \\
                                    \\
$B_{\perp}$           &$0.10^{+0.40+0.10+0.03}_{-0.07-0.02-0.00}$          &$0.01^{+0.04+0.01+0.00}_{-0.01-0.00-0.00}$            \\
                                  \\
$B_{total}$           &$0.90^{+0.69+0.37+0.12}_{-0.30-0.23-0.04}$           &$0.41^{+0.17+0.13+0.02}_{-0.13-0.12-0.01}$            \\
\\
\hline
\hline
\end{tabular*}
\end{table}

In order to compare the branching ratios with other approaches and experimental results of two body decay~\cite{Yan:2018ag,Ali:2007co,Zou:2015de,Cheng:2009df,Wang:2017bf,Particle Date Group:2020ac}, we predict branching ratios of the corresponding two body vector resonance by the following relation Eq.~(\ref{equ:abc}) and ${\cal B}(\rho^{0}\rightarrow \pi\pi)=1$.

As aforementioned, the relation of branching ratios between two body vector resonance and corresponding quasi-two body decay in narrow-width approximation has been obtained as

\begin{equation}\label{equ:abc}
{\cal B}(B^{0}_{s} \rightarrow \rho^{0}(\rightarrow \pi\pi) \rho^{0}(\rightarrow \pi\pi))\thickapprox {\cal B}(B^{0}_{s} \rightarrow \rho^{0}\rho^{0})\times {\cal B}(\rho^{0}\rightarrow \pi\pi)\times {\cal B}(\rho^{0}\rightarrow \pi\pi).
\end{equation}

The CP-averaged branching ratio which we predicted for $B^{0}_{S}\rightarrow\rho^{0}\rho^{0}$ is $0.41\times10^{-8}$ while for $B^{0}_{S}\rightarrow\rho^{+}\rho^{-}$ is $0.90\times10^{-8}$. Our branching ratios are in agreement with the results of FAT approaches and QCDF approaches within errors, however, they are lower than the results of previous PQCD~\cite{Zou:2015de,Cheng:2009df}. The branching ratio of $B^{0}_{S}\rightarrow\rho^{0}\rho^{0}$ is lower than the upper bound in experiment, all other annihilation decays have not been measured and are excepted to be confirmed by future experiments.

We now discuss the predictions for the polarization fraction of $B^{0}_{S}$ meson.
For pure annihilation two-body decays, the contributions are dominated by the longitudinal polarization fraction $f_{0}$ and the fractions of these decays can reach to about 100\%, which have been pointed in previous predictions of two-body decays. It is found that our results are in agreement well with former fractions~\cite{Zou:2015de,Cheng:2009df}, so we predict polarization fraction in four-body decays mainly in this paper.

As shown in the following, we can find that for decay $B^{0}_{S}\rightarrow(\rho^{+}\rightarrow)\pi^{+}\pi^{0}(\rho^{-}\rightarrow)\pi^{-}\pi^{0}$, the longitudinal polarization fraction $f_{0}$ is about 84.44\%, and $f_{0}$ is about 92.68\% for $B^{0}_{S}\rightarrow(\rho^{0}\rightarrow)\pi^{+}\pi^{-}(\rho^{0}\rightarrow)\pi^{+}\pi^{-}$ decay. The uncertainties of results come from QCD scale and the hard scale. The results show that the transverse polarization cannot be ignored and can help to make significant contributions in pure annihilation decays.
\begin{eqnarray}
f_{0}=
\left\{
  \begin{array}{lll}
  &84.44^{+0.24+8.20}_{-0.06-2.82}\% \qquad B^{0}_{S}\rightarrow(\rho^{+}\rightarrow)\pi^{+}\pi^{0}(\rho^{-}\rightarrow)\pi^{-}\pi^{0},                           \\
                                                                                     \\
 &92.68^{+1.60+3.18}_{-0.08-1.50}\% \qquad B^{0}_{S}\rightarrow(\rho^{0}\rightarrow)\pi^{+}\pi^{-}(\rho^{0}\rightarrow)\pi^{+}\pi^{-}                           .
  \end{array}
 \right.
\end{eqnarray}

\begin{table}[htbp]
\centering
\caption{The CP-averaged branching ratios of the $B^{0}_{S} \rightarrow N_{1}N_{2} \rightarrow \pi\pi\pi\pi$ decay, uncertainties of these results come from the shape parameter, hard scale and Gegenbauer moments.}
\label{6666}
\begin{tabular*}{\columnwidth}{@{\extracolsep{\fill}}lllll@{}}
\hline
\hline
Modes     &${\cal B}(10^{-8})$       \\
\hline
\\
$B^{0}_{S}\rightarrow(f_{0}(980)\rightarrow)\pi^{+}\pi^{-}(f_{0}(980)\rightarrow)\pi^{+}\pi^{-}$                 &$11.75^{+2.55+0.04+0.00}_{-2.89-1.60-0.03}$   \\
                                   \\
$B^{0}_{S}\rightarrow(\rho^{0}\rightarrow)\pi^{+}\pi^{-}(f_{0}(500)\rightarrow)\pi^{+}\pi^{-}$            &$0.11^{+0.05+0.05+0.00}_{-0.02-0.01-0.00}$  \\
\\
$B^{0}_{S}\rightarrow(f_{0}(980)\rightarrow)\pi^{+}\pi^{-}(f_{0}(980)\rightarrow)\pi^{0}\pi^{0}$                   &$5.88^{+1.28+0.02+0.00}_{-1.45-0.80-0.01}$   \\
                                   \\
$B^{0}_{S}\rightarrow(\rho^{0}\rightarrow)\pi^{+}\pi^{-}(f_{0}(500)\rightarrow)\pi^{0}\pi^{0}$            &$0.06^{+0.02+0.02+0.00}_{-0.01-0.01-0.00}$  \\
\\
$B^{0}_{S}\rightarrow(f_{0}(980)\rightarrow)\pi^{0}\pi^{0}(f_{0}(980)\rightarrow)\pi^{0}\pi^{0}$                   &$2.94^{+0.64+0.01+0.00}_{-0.72-0.40-0.01}$   \\
\hline
\hline
\end{tabular*}
\end{table}

Compared with the decays of double vector resonant states, the decays of scalar resonances, which have less experimental data, are more difficult to predict because of the large decay width in scalar resonances. In Table \ref{6666}, we calculate the four body decays of $B^{0}_{S}\rightarrow(f_{0}(980)\rightarrow)\pi\pi(f_{0}(980)\rightarrow)\pi\pi$ and $B^{0}_{S}\rightarrow(\rho^{0}\rightarrow)\pi\pi(f_{0}(500)\rightarrow)\pi\pi$, we ignore the decay of $B^{0}_{S}\rightarrow(f_{0}(500)\rightarrow)\pi\pi(f_{0}(500)\rightarrow)\pi\pi$ because of its lower branching ratio. Comparing Table \ref{6666} with Table \ref{4444}, we find that the decay of $B^{0}_{S}\rightarrow(f_{0}(980)\rightarrow)\pi^{+}\pi^{-}(f_{0}(980)\rightarrow)\pi^{+}\pi^{-}$ is the largest contribution in total branching ratio, because there is only pure annihilation contribution in decay of $B^{0}_{S}\rightarrow\rho\rho\rightarrow \pi\pi\pi\pi$. This results have not been reported by experiments and are expected to be studied in future LHCb and Belle-II experiments.

\begin{table}[htbp]
\centering
\caption{The CP-violating asymmetries of the $B^{0}_{S} \rightarrow N_{1}N_{2} \rightarrow \pi\pi\pi\pi$ decay, the errors come from QCD scale, the Gegenbauer moments and hard scale.}
\label{7777}
\begin{tabular*}{\columnwidth}{@{\extracolsep{\fill}}lllll@{}}
\hline
\hline
$Asymmetries$ &$B^{0}_{S}\rightarrow(\rho^{+}\rightarrow)\pi^{+}\pi^{0}(\rho^{-}\rightarrow)\pi^{-}\pi^{0}$ &$B^{0}_{S}\rightarrow(\rho^{0}\rightarrow)\pi^{+}\pi^{-}(\rho^{0}\rightarrow)\pi^{+}\pi^{-}$ \\
\hline
\\
$A^{dir}$   &$(6.71^{+0.00+0.00+0.00}_{-3.42-4.71-5.43})\%$        &$(7.22^{+0.25+1.84+1.93}_{-0.00-0.38-1.02})\%$              \\
\\
$A^{dir}_{0}$    &$(7.64^{+0.00+0.00+0.00}_{-3.62-3.66-6.32})\%$        &$(7.68^{+0.00+0.00+1.77}_{-0.00-4.02-5.14})\%$   \\
\\
$A^{dir}_{\parallel}$   &$(0.02^{+1.28+6.31+6.50}_{-0.00-0.00-0.00})\%$  &$(0.05^{+0.40+2.49+2.66}_{-0.00-0.00-0.00})\%$   \\
\\
$A^{dir}_{\perp}$    &$(1.66^{+0.00+0.00+5.55}_{-0.25-0.50-1.66})\%$ &$(3.28^{+0.00+0.00+0.00}_{-0.50-0.99-3.28})\%$ \\
\\
\hline
$Asymmetries$ &$B^{0}_{S}\rightarrow(\rho^{0}\rightarrow)\pi^{+}\pi^{-}(f_{0}(500)\rightarrow)\pi^{+}\pi^{-}$ &$B^{0}_{S}\rightarrow(f_{0}(500)\rightarrow)\pi^{+}\pi^{-}(f_{0}(500)\rightarrow)\pi^{+}\pi^{-}$ \\
\hline
                                  \\
$A^{dir}$   &$(11.92^{+0.00+1.94+2.17}_{-0.70-1.43-3.68})\%$  &$(0.26^{+0.46+1.42+12.34}_{-0.00-0.00-0.04})\%$    \\
\\
\hline
$Asymmetries$ &$B^{0}_{S}\rightarrow(f_{0}(980)\rightarrow)\pi^{+}\pi^{-}(f_{0}(980)\rightarrow)\pi^{+}\pi^{-}$   \\
\hline
\\
$A^{dir}$                &0.00\%        \\
\hline
\hline
\end{tabular*}
\end{table}

For overall direct CP asymmetry, we define

\begin{equation}
A^{dir}=\frac{\overline{{\cal B}}_{total}-{\cal B}_{total}}{\overline{{\cal B}}_{total}+{\cal B}_{total}},
\end{equation}
where

\begin{equation}
{\cal B}_{total}={\cal B}_{0}+{\cal B}_{\parallel}+{\cal B}_{\perp}.
\end{equation}

The direct CP asymmetry in each component can be defined as

\begin{equation}
A^{dir}_{h}=\frac{\overline{{\cal B}}_{h}-{\cal B}_{h}}{\overline{{\cal B}}_{h}+{\cal B}_{h}},
\end{equation}
here $h=0, \parallel, \perp$.

The CP-violating asymmetries are listed in Table \ref{7777}. For $B^{0}_{S}\rightarrow(f_{0}(980)\rightarrow)\pi^{+}\pi^{-}(f_{0}(980)\rightarrow)\pi^{+}\pi^{-}$ decay, because it is pure penguin process with transition $b \rightarrow ss\overline{s}$, the result is small or even zero. For pure annihilation type decay process, the CP-violating asymmetries are also small. The results has been discussed by previous works~\cite{Zou:2015de,Cheng:2009df}.
However, we found that CP-violating asymmetry can be enhanced largely by the $\rho-\omega$ mixing resonances when $\pi\pi$ pairs masses are in the vicinity of $\omega$ resonance~\cite{Lu:2013dg}, so it is important for us to study CP-violating asymmetry via $\rho$ and $\omega$ resonances in three-body and four-body decays. The result of $B^{0}_{S}\rightarrow\rho^{0}(\omega)\rho^{0}(\omega)\rightarrow\pi^{+}\pi^{-}\pi^{+}\pi^{-}$ are listed in Table \ref{8888}. We also compare our prediction with previous result~\cite{Lu:2017dh}, as we can see, for the unpolarized CP-violating asymmetry, our result is in agreement with previous result, however, our result has big errors because of the different approach we adopted.

\begin{table}[htbp]
\centering
\caption{The CP-violating asymmetries of the $B^{0}_{S} \rightarrow \rho^{0}(\omega)\rho^{0}(\omega)\rightarrow \pi^{+}\pi^{-}\pi^{+}\pi^{-}$ decay, the errors come from the Gegenbauer moments, hard scale and QCD scale.}
\label{8888}
\begin{tabular*}{\columnwidth}{@{\extracolsep{\fill}}lllll@{}}
\hline
\hline
$Asymmetries$ &$B^{0}_{S}\rightarrow\rho^{0}(\omega)\omega(\rho^{0})\rightarrow\pi^{+}\pi^{-}\pi^{+}\pi^{-}$(this work) &$B^{0}_{S}\rightarrow\rho^{0}(\omega)\rho^{0}(\omega)\rightarrow\pi^{+}\pi^{-}\pi^{+}\pi^{-}$(~\cite{Lu:2017dh}) \\
\hline
\\
$A^{dir}$ &$(27.30^{+8.04+8.38+8.60}_{-0.00-17.78-20.71})\%$  &$(27.20^{+0.05+0.28+7.13}_{-0.15-0.31-6.11})\%$          \\
\\
$A^{dir}_{0}$ &$(37.02^{+0.00+8.30+14.25}_{-0.01-24.82-26.98})\%$   &... \\
\\
$A^{dir}_{\parallel}$ &$(0.14^{+3.01+3.08+3.55}_{-0.00-0.00-0.00})\%$ &... \\
\\
$A^{dir}_{\perp}$  &$(11.56^{+0.00+0.12+1.96}_{-0.66-0.92-10.22})\%$  &... \\
\hline
\hline
\end{tabular*}
\end{table}

\section{Summary} \label{sec:summary}

In this work, we study the CP-averaged branching ratios and direct CP-violating asymmetries of the quasi-two-body decays $B^{0}_{S} \rightarrow N_{1}N_{2} \rightarrow \pi\pi\pi\pi$ decay from the S-wave resonances, $f_{0}(980)$ and $f_{0}(500)$ and P-wave resonances, $\rho(770)$, by introducing the S-wave and P-wave $\pi\pi$ distribution amplitudes within the framework of the perturbative QCD approach. We also calculate branching ratios of the two-body decays $B^{0}_{S}\rightarrow\rho^{0}\rho^{0}$, $B^{0}_{S}\rightarrow\rho^{+}\rho^{-}$ from the corresponding quasi-two-body decays models and compare our results with those obtained in previous perturbative QCD approach, QCD factorization approach and FAT approach. The predictions are in agreement with present data within errors. For the CP-violating asymmetries, it is small in pure annihilation type decay process, however, we found that CP-violating asymmetry can be enhanced largely by the $\rho-\omega$ mixing resonances when $\pi\pi$ pairs masses are in the vicinity of $\omega$ resonance.


\section*{acknowledgments}

The authors would to thank Dr. Ming-Zhen Zhou for some valuable discussions. This work is supported by the National Natural Science Foundation of China under Grant No.11047028.

\section*{Appendix : Formulae For The Calculation Used In The Text} \label{sec:appendix}
\appendix
\setcounter{equation}{0}
\renewcommand\theequation{A.\arabic{equation}}
In this section, we list the decay amplitude for each considered decay mode of four body B meson.

\begin{equation}
\begin{split}
A_{h}(B^{0}_{s} \rightarrow (\rho^{+}\rightarrow)\pi^{+}\pi^{0}(\rho^{-}\rightarrow)\pi^{-}\pi^{0})
&=\frac{G_{F}}{\sqrt{2}}({V^{*}_{ub}V_{us}[(C_{1}+\frac{1}{3}C_{2})F^{ll,h}_{a}+C_{2}M^{ll,h}_{a}]}\\
&-{2}V^{*}_{tb}V_{ts}[({{2}C_{3}+\frac{2}{3}C_{4}+{2}C_{5}+\frac{2}{3}C_{6}}\\
&+{\frac{1}{2}C_{7}+\frac{1}{6}C_{8}+\frac{1}{2}C_{9}+\frac{1}{6}C_{10}})F^{ll,h}_{a}\\
&+({2}C_{4}+\frac{1}{2}C_{10})M^{ll,h}_{a}+({2}C_{6}+\frac{1}{2}C_{8})M^{sp,h}_{a}]),
\end{split}
\end{equation}

\begin{equation}
\begin{split}
A_{h}(B^{0}_{s} \rightarrow (\rho^{0}\rightarrow )\pi\pi(\rho^{0}\rightarrow )\pi\pi)&={G_{F}}({V^{*}_{ub}V_{us}[(C_{1}+\frac{1}{3}C_{2})F^{ll,h}_{a}+C_{2}M^{ll,h}_{a}]}\\
&-V^{*}_{tb}V_{ts}[({{2}C_{3}+\frac{2}{3}C_{4}+{2}C_{5}+\frac{2}{3}C_{6}}\\
&+{\frac{1}{2}C_{7}+\frac{1}{6}C_{8}+\frac{1}{2}C_{9}+\frac{1}{6}C_{10}})F^{ll,h}_{a}\\
&+({2}C_{4}+\frac{1}{2}C_{10})M^{ll,h}_{a}+({2}C_{6}+\frac{1}{2}C_{8})M^{sp,h}_{a}]),
\end{split}
\end{equation}

\begin{equation}
\begin{split}
A_{h}(B^{0}_{s} \rightarrow (\rho^{0}\rightarrow )\pi\pi(\omega\rightarrow )\pi\pi)&=\frac{G_{F}}{\sqrt{2}}({V^{*}_{ub}V_{us}[(C_{1}+\frac{1}{3}C_{2})F^{ll,h}_{a}+C_{2}M^{ll,h}_{a}]}\\
&-V^{*}_{tb}V_{ts}[({\frac{3}{2}C_{7}+\frac{1}{2}C_{8}}
+{\frac{3}{2}C_{9}+\frac{1}{2}C_{10}})F^{ll,h}_{a}\\
&+(\frac{3}{2}C_{10})M^{ll,h}_{a}+(\frac{3}{2}C_{8})M^{sp,h}_{a}]).
\end{split}
\end{equation}

Here $G_{F}=1.16639\times10^{-5} {\rm GeV}^{-2}$ is the Fermi coupling constant. For the double P-wave resonance, we decompose the decay amplitudes into three helicity components with $h=0, \parallel, \perp.$ $(V-A)\otimes(V-A)$, $(V-A)\otimes(V+A)$, $(S-P)\otimes(S+P)$ are defined as LL, LR and SP. $F_{e}$ and $M_{e}$ refer to the factorizable or nonfactorizable emission diagrams, $F_{a}$ and $M_{a}$ refer to the factorizable or nonfactorizable annihilation diagrams.

\begin{equation}
\begin{split}
A(B^{0}_{s} \rightarrow (\rho^{0}\rightarrow )\pi\pi(f_{0}(500)\rightarrow )\pi\pi)&=\frac{G_{F}}{\sqrt{2}}({V^{*}_{ub}V_{us}[(C_{1}+\frac{1}{3}C_{2})F^{ll,vs}_{a}+C_{2}M^{ll,vs}_{a}]}\\
&-V^{*}_{tb}V_{ts}[({\frac{3}{2}C_{7}+\frac{1}{6}C_{8}}
+{\frac{3}{2}C_{9}+\frac{1}{2}C_{10}})F^{ll,vs}_{a}\\
&+(\frac{3}{2}C_{10})M^{ll,vs}_{a}+(\frac{3}{2}C_{8})M^{sp,vs}_{a}]),
\end{split}
\end{equation}

\begin{equation}
\begin{split}
A(B^{0}_{s} \rightarrow (f_{0}(980)\rightarrow )\pi\pi(f_{0}(980)\rightarrow )\pi\pi)
&=-{\sqrt{2}}{G_{F}}V^{*}_{tb}V_{ts}[({\frac{4}{3}C_{3}+\frac{4}{3}C_{4}+C_{5}+\frac{1}{3}C_{6}}\\
&{-\frac{1}{2}C_{7}-\frac{1}{6}C_{8}-\frac{2}{3}C_{9}-\frac{2}{3}C_{10}})F^{ll,ss}_{a}\\
&+{(C_{6}+\frac{1}{3}C_{5}-\frac{1}{2}C_{8}-\frac{1}{6}C_{7})(F^{sp,ss}_{e}+F^{sp,ss}_{a})}\\
&+{(C_{3}+C_{4}-\frac{1}{2}C_{9}-\frac{1}{2}C_{10})(M^{ll,ss}_{e}+M^{ll,ss}_{a})}\\
&+{(C_{5}-\frac{1}{2}C_{7})(M^{lr,ss}_{e}+M^{lr,ss}_{a})}\\
&+{(C_{6}-\frac{1}{2}C_{8})(M^{sp,ss}_{e}+M^{sp,ss}_{a})}].
\end{split}
\end{equation}

Here a S-wave resonance is described as $vs$, and $ss$ stands for the double S-wave resonance component. The explicit expressions for the the factorizable contributions $F_{a,e}$ and the nonfactorizable contributions $M_{a,e}$ from Fig.~\ref{fig:fig2} can be found in Refs~\cite{Rui:2021bw,Li:2021bx,Zhang:2021ch}.


\end{document}